\documentclass[10pt]{article}
\usepackage{epsfig,amsmath,graphicx,amssymb,overpic}
\def\be{\begin{equation}}
\def\ee{\end{equation}}
\def\bee{\begin{eqnarray}}
\def\ene{\end{eqnarray}}
\def\bes{\begin{subequations}}
\def\ees{\end{subequations}}

\newcommand{\bR}{{\rm R}}
\newcommand{\bI}{{\rm I}}

\begin{document}

\baselineskip=13pt
\renewcommand {\thefootnote}{\dag}
\renewcommand {\thefootnote}{\ddag}
\renewcommand {\thefootnote}{ }

\pagestyle{plain}

\begin{center}
\baselineskip=16pt \leftline{} \vspace{-.3in} {\large \bf
Nonautonomous ``rogons" in the inhomogeneous nonlinear
Schr\"odinger equation

with variable coefficients
} \\[0.2in]
\end{center}

\begin{center}

{ Zhenya Yan$^{\rm a,b}$}\footnote{{\it Email address}: zyyan@mmrc.iss.ac.cn} \\[0.03in]
{\it \small $^{\rm a}$Centro de F\'isica Te\'orica e
Computacional, Universidade de Lisboa,  Complexo Interdisciplinar,
\\ Lisboa
1649-003, Portugal \\
 $^{\rm b}$Key Laboratory of Mathematics Mechanization, Institute
of Systems Science, AMSS,  \\ Chinese Academy of Sciences, Beijing
100190, China}

\end{center}

\vspace{0.1in}

{\baselineskip=19pt

\begin{tabular}{p{16cm}}
 \hline \\
\end{tabular}

\vspace{-0.18in}

\begin{abstract} \small \baselineskip=14pt

The analytical nonautonomous rogons are reported for the
inhomogeneous nonlinear Schr\"odinger equation with variable
coefficients in terms of rational-like functions by using the
similarity transformation and direct ansatz. These obtained
solutions can be used to describe the possible formation
mechanisms for optical, oceanic, and matter rogue wave phenomenon
in optical fibres, the deep ocean, and Bose-Einstein condensates,
respectively. Moreover, the snake propagation traces and the
fascinating interactions of two nonautonomous rogons are generated
for the chosen different parameters. The obtained nonautonomous
rogons may excite the possibility of relative experiments and
potential applications for the rogue wave phenomenon in the field
of nonlinear science.

\vspace{0.1in} \noindent {\it Key words:}  \ Inhomogeneous NLS
equation with variable coefficients; Similarity transformation;

\qquad\qquad Rational-like solutions; Rogue waves; Rogons

\vspace{0.1in} \noindent {\it PACS:} \ 05.45.Yv; 42.65.-k;
42.81.Dp; 42.65.Sf; 03.75.Lm

\end{abstract}

\vspace{-0.05in}
\begin{tabular}{p{16cm}}
  \hline \\
\end{tabular}

\vspace{-0.15in}

\vspace{0.5in}




\quad The nonlinear Schr\"odinger (NLS) equation is a foundational
model describing numerous nonlinear physical phenomenon in the field
of nonlinear science such as optical solitons in optical
fibres~\cite{Sulem, Opt1, Opt2}, solitons in the mean-field theory
of Bose-Einstein condensates~\cite{bec1,bec2,bec3}, and the {\it
rogue waves} (RWs) (also known as {\it freak waves}, {\it monster
waves}, {\it killer waves}, {\it giant waves} or {\it extreme
waves}) in the nonlinear oceanography~\cite{Ocean, Ocean1,
Ocean11,Ocean12}. RWs are single waves generated in the ocean with
amplitudes much higher than the average wave crests around
them~\cite{RW, RW2}. Recently, the RWs have attracted more and more
attention from the point views of both theoretical
analysis~\cite{ABC, ABC1, ABC2, couple, couple2, ABCP, ABCP2, BRW}
and experimental realization~\cite{Ocean2, Ocean22, exp00, exp0,
exp1, exp2}. The oceanic RWs can be, under the nonlinear theories of
ocean waves, modelled by the dimensionless NLS equation~\cite{Ocean,
Ocean1}
 \bee
  \label{nls}
   i\frac{\partial\psi}{\partial t}+\frac{1}{2}\frac{\partial^2\psi}{\partial x^2}+|\psi|^2\psi=0,
   \qquad\qquad\qquad\qquad\qquad\qquad\qquad\qquad\qquad\qquad \ene
which describes the two-dimensional quasi-periodic deep-water
trains in the lowest order in wave steepness and spectral width.
In addition, it has been shown that the RWs can be generated in
nonlinear optical systems and the term ``optical rogue waves" was
coined by observing optical pulse propagation in the generalized
NLS equaiton~\cite{Ocean2, Ocean22}. Some types of exact solutions
of Eq. (\ref{nls}) have been presented to describe the possible
formation mechanisms for the RW phenomenon such as the algebraic
breathers (Peregrine solitons)~\cite{AB}, the time periodic
breather (Ma solitons)~\cite{TP}, the space periodic breathers
(Akhmediev breathers)~\cite{SP, SP2}. More recently, the
 Akhmediev breathers were also
further studied~\cite{ABS, ABS2, ABC, ABC1}. A possible mechanism
for the formation of RWs was also exhibited by using
two-dimensional coupled NLS equations ~\cite{couple,
couple2,couple3} describing the nonlinearly interacting
two-dimensional waves in deep water. The three-dimensional
mechanism of RW formation was studied in a late stage of the
modulational instability of a perturbed Stokes deep-water
wave~\cite{rw3}. Furthermore, the optical RWs were also found in
the NLS equation with perturbing terms (the higher-order NLS
equation with the third-order dispersion, self-steeping, and
self-frequency shift) ~\cite{ABC2}. Recently, the existence of
matter RWs in Bose-Einstein condensates was predicted to either
load into a parabolic trap or embed in an optical
lattice~\cite{BRW}.

\quad Here we point out that based on physical similarities between
the rogue waves and solitary waves, first observed by Russell in
1834 and further known as solitons by Zabusky and Kruskal in
1965~\cite{soliton}, we coin the  word ``Rogon" for each such
``Rogue Wave"  (or the word ``Freakon" for the ``Freak Wave") if
they reappear virtually unaffected in size or shape shortly after
their interactions. Similarly, there also exist the corresponding
new terms ``oceanic rogons", ``optical rogons", and ``matter rogons"
for the oceanic rogue waves~\cite{Ocean, Ocean1, Ocean11, Ocean12},
optical rogue waves~\cite{Ocean2, Ocean22, ABC, ABC1, ABC2, ABCP,
ABCP2}, and matter rogue waves~\cite{BRW}, respectively.

\quad To the best of our knowledge, there was no report on exact
solutions related to the modified RWs (rogons) with variable
functions before. In this Letter, we will extend the NLS equation
(\ref{nls}) to the inhomogeneous NLS equation with variable
coefficients, including group-velocity dispersion $\beta(t)$,
linear potential $v(x,t)$, nonlinearity $g(t)$ and the gain/loss
term $\gamma(t)$, in the form~\cite{1DNLS, 1DNLS2, 1DNLS3, 1DNLS4}
\bee \label{nlsv} i\frac{\partial\psi}{\partial
t}+\frac{\beta(t)}{2}\frac{\partial^2\psi}{\partial
x^2}+v(x,t)\psi+ g(t)|\psi|^2\psi=i\gamma(t)\psi,
\qquad\qquad\qquad\qquad\qquad\qquad \ene which is associated with
$\delta\mathcal{L}/\delta \psi^*=0$, in which the Lagrangian
density is written as $
 \mathcal{L}=i(\psi_t\psi^*-\psi\psi^*_t)-\beta(t)|\psi_x|^2
+g(t)|\psi|^4+2[v(x,t)-i\gamma(t)]|\psi|^2$, where
$\psi\equiv\psi(x,t)$, and $\psi^*$ denotes the complex conjugate
of the physical field $\psi$. Eq. (\ref{nlsv}) can be also known
as the generalized Gross-Pitaevskii equation with variable
coefficients for $\beta(t)=1$~\cite{bec1,bec2,bec3,yank}. Our goal
is focused on the first-order and second-order rational-like
solutions of Eq. (\ref{nlsv}) to describe the possible formation
mechanisms of optical RWs by using the similarity
transformations~\cite{simi, simi1, simi2, simi3, simi4, yank} and
direct ansatz~\cite{ABC, ABC1}. Moreover, we analyze the dynamical
behaviors of these solutions and interactions of two optical RWs
(rogons) by choosing the different functions.


\quad {\it Similarity transformation and nonautonomous rogons}---
To investigate the analytical rational-like solutions of
Eq.(\ref{nlsv}) related to the optical nonautonomous rogons, we
employ the envelope field $\psi(x, t)$ in the gauge
form~\cite{simi2, simi3, simi4,yank}
 \bee
\label{comtran}
\begin{array}{l}
 \psi(x,t)=\left[\Psi_{\bR}(x, t)+i\Psi_\bI(x,t)\right]e^{i\varphi(x,t)}
 \qquad\qquad\qquad\qquad\qquad\qquad\qquad\qquad \end{array} \ene
whose intensity can be written as $|\psi(x,t)|^2=|\Psi_\bR(x,
t)|^2+|\Psi_\bI(x, t)|^2$, where $\Psi_\bR(x, t),\ \Psi_\bI(x,t)$
and $\varphi(x,t)$ are real functions of space-time $x,t$. The
substitution of transformation (\ref{comtran}) into Eq.
(\ref{nlsv}) yields the system of coupled real partial
differential equations with variable coefficients
\bes \label{sysc}
 \bee \label{sysc1}  &&
 \Psi_{\bR,t}+\frac{\beta(t)}{2}\left(\Psi_{\bI,xx}+2\varphi_x\Psi_{\bR,x}-\varphi_x^2\Psi_\bI
 +\varphi_{xx}\Psi_\bR\right)+[v(x,t)-\varphi_t]\Psi_\bI+g(t)(\Psi_\bR^2+\Psi_\bI^2)\Psi_\bI-\gamma(t)\Psi_\bR=0,
 \qquad
  \vspace{0.1in} \\
\label{sysc2} &&
-\Psi_{\bI,t}+\frac{\beta(t)}{2}\left(\Psi_{\bR,xx}-2\varphi_x\Psi_{\bI,x}-\varphi_x^2\Psi_\bR
  -\varphi_{xx}\Psi_\bI\right)+[v(x,t)-\varphi_t]\Psi_\bR+g(t)(\Psi_\bR^2+\Psi_\bI^2)\Psi_\bR+\gamma(t)\Psi_\bI=0.
  \qquad
  \ene \ees

\quad By introducing the new variables $\eta(x,t)$ and $\tau(t)$,
we further utilize the following similarity transformations for
the real functions $\Psi_\bR(x,t)$, $\Psi_\bI(x,t)$ and the phase
$\varphi(x,t)$
 \bee
 \label{newtran} \begin{array}{l}
 \Psi_\bR(x,t)=A(t)+B(t)P(\eta(x,t),\tau(t)), \qquad\qquad\qquad\qquad\qquad\qquad\qquad\qquad \vspace{0.1in} \cr
 \Psi_\bI(x,t)=C(t)Q(\eta(x,t),\tau(t)), \vspace{0.1in} \cr
 \varphi(x,t)=\chi(x,t)+\mu\tau(t),
 \end{array}
 \ene
 to system (\ref{sysc}) such that we deduce the following similarity reduction
 \bes \vspace{-0.15in}
\label{sys} \bee
\label{sys1} && \eta_{xx}=0, \\
\label{sys2} &&\eta_t+\beta(t)\chi_x\eta_x=0, \\
\label{sys3} && 2\chi_t+\beta(t)\chi_x^2-2v(x,t)=0, \\
\label{sys4} && 2\sigma_t+[\beta(t)\chi_{xx}-2\gamma(t)] \sigma=0  \  \ \ (\sigma=A,B,C),\\
\label{sys5} && \tau_tBP_{\tau}+\frac{\beta(t)}{2}\eta_x^2CQ_{\eta\eta}-\mu \tau_tCQ +g(t)CQ[C^2Q^2+(A+BP)^2]=0, \\
\label{sys6}  &&
-\tau_tCQ_{\tau}+\frac{\beta(t)}{2}\eta_x^2BP_{\eta\eta}-\mu\tau_t
(A+BP) +g(t)(A+BP)[C^2Q^2+(A+BP)^2]=0, \qquad\qquad \ene \ees
where $\mu$ is a constant, and $\eta(x, t),\  \chi(x, t),\ A(t),\
B(t), \ C(t),\ \tau(t),\ P(\eta,\tau),\ Q(\eta,\tau)$ are
functions to be determined. After some algebra, it follows from
Eqs. (\ref{sys1})-(\ref{sys4}) that we have \bes
 \label{simis}
 \bee
 \label{simis1} && \eta(x,t)=\alpha(t)x+\delta(t), \quad
 \chi(x,t)=-\frac{\alpha_t}{2\alpha(t)\beta(t)}x^2-\frac{\delta_t}{\alpha(t)\beta(t)}x+\chi_0(t), \quad
 v(x,t)=\chi_t+\frac{\beta(t)}{2}\chi_x^2, \qquad \\
 && \label{simis2} A(t)=a_0\sqrt{|\alpha(t)|}\,e^{\int^t_0\gamma(s) ds},
  \qquad B(t)=bA(t),\  \ \ \  C(t)=cA(t),
\ene \ees where $a_0, \ b,\ c$ are constants,   $\alpha(t)$ (the
inverse of the wave width),\ $\delta(t)$ ($-\delta(t)/\alpha(t)$
being the position of its center of mass),
 and $\chi_0(t)$ are all free functions of time $t$.

\quad To further reduce Eqs. (\ref{sys5}) and (\ref{sys6}) to the
system of coupled partial differential equations with constant
coefficients, We require the conditions:
$\tau_t=\frac{\beta(t)}{2}\eta_x^2$ and $
  g(t)=\frac{\beta(t)}{2}GA^{-2}(t)\eta_x^2 \ (G={\rm const.})$, which can generate the
  constraints for the variable $\tau(t)$ and nonlinearity $g(t)$
 \bee \label{solu3}
 \tau(t)=\frac{1}{2}\int_0^t \alpha^2(s)\beta(s) ds, \qquad\qquad
 g(t)=\frac{G\alpha(t)\beta(t)}{2a_0^2e^{2\int^t_0\gamma(s) ds}} \qquad\qquad\qquad\qquad\qquad
 \ene
such that Eqs. (\ref{sys5}) and (\ref{sys6}) reduce to the coupled
system of differential equation with constant coefficients
  \bes \label{syst}
\bee
 && bP_{\tau}+cQ_{\eta\eta}-\mu cQ+cGQ[c^2Q^2+(1+bP)^2]=0,  \\
\label{syst2}  &&
-cQ_{\tau}+bP_{\eta\eta}-\mu(1+bP)+G(1+bP)[c^2Q^2+(1+bP)^2]=0.
\qquad\qquad\qquad\qquad\qquad \ene \ees


\quad Following the direct approach developed in Refs. \cite{ABCP,
ABCP2}, we can obtain the rational solutions of system
(\ref{syst}) such that the corresponding rational-like solutions
(nonautonomous rogons) of Eq. (\ref{nlsv}) can be found in terms
of similarity transformations (\ref{comtran})  and
(\ref{newtran}). In the following, we will exhibit the dynamical
behaviors of rational-like solutions with many interesting
nontrivial features.

\quad {\it First-order rational-like solution--} It follows from
system (\ref{syst}) that we have the solution
$P(\eta,\tau)=-4/[b\mathcal{A}_1(\eta,\tau)]$ and $
Q(\eta,\tau)=-8\tau/[c\mathcal{A}_1(\eta,\tau)]$ with
$\mathcal{A}_1(\eta,\tau)=1+2\eta^2+4\tau^2$ for $\mu=G=1$. Thus
based on the similarity transformations (\ref{comtran}) and
(\ref{newtran}), we obtain the first-order rational-like solution
(nonautonomous rogon) of Eq. (\ref{nlsv})
 \bee
\label{solu1}
\psi_1(x,t)=a_0\sqrt{|\alpha(t)|}\,e^{\int^t_0\gamma(s)ds}
 \left[1-\frac{4+8i\tau(t)}{1+2[\alpha(t)x+\delta(t)]^2+4\tau^2(t)}\right]e^{i[\chi(x,t)+\tau(t)]},
\ene whose intensity can be written as
  \bee \label{inten1}
   |\psi_1(x,t)|^2=a_0^2|\alpha(t)|\,e^{2\int^t_0\gamma(s)ds}\frac{\left\{2[\alpha(t)x+\delta(t)]^2
   +4\tau^2(t)-3\right\}^2+64\tau^2(t)}
   {\left\{1+2[\alpha(t)x+\delta(t)]^2+4\tau^2(t)\right\}^2},  \qquad \ene
where $\chi(x,t)$ and $\tau(t)$ are given by Eqs. (\ref{simis1})
and (\ref{solu3}).

 \quad It is easy to see that the rational-like
solution (\ref{solu1}) is different from the known rational
 solution of the NLS equation (\ref{nls})~\cite{AB, TP, ABC, ABC1, ABC2,
 ABCP, ABCP2}, since it contains some free functions of time $t$, which
 will generate abundant structures related to the optical RWs. In particular, when $\alpha=2,\ a_0=\beta=1,\
 \chi_0=\gamma=0$, resulting in $g=1$, Eq. (\ref{nlsv}) reduces to Eq. (\ref{nls}) such
 that the nonautonomous rogon solution (\ref{solu1}) reduces to the known rogon solution in Refs.~\cite{ABC,ABCP}.
 In what follows, we will choose some free functions of time to exhibit the obtained  rational-like solution (\ref{solu1}).

\quad  For the fixed parameters $\alpha_0=1, \ \chi_0(t)=0$ and
$\gamma(t)=0.1\tanh(t){\rm sech}(t)$, i) if we choose other free
functions as
 the polynomials of time $t$, i.e. $\alpha(t)=1, \ \beta(t)=0.5t^2$,
 then Figures 1a and 1b depict the dynamical behavior of the rational-like solution (\ref{solu1}) for different
 terms $\delta(t)=t,\ t^2$, respectively, in which the other coefficients $g(t)$ and $v(x,t)$ in Eq. (\ref{nlsv})
 are given by
 \bes
 \bee
 && g(t)=\frac14t^2e^{[{\rm sech}(t)-1]/5}, \\
 && v(x,t)=4t^{-3}x+t^{-2} \ \ \ {\rm for} \ \ \ \delta(t)=t, \\
 && v(x,t)=4t^{-2}x+4 \ \ \ {\rm for} \ \ \ \delta(t)=t^2; \qquad\qquad\qquad\qquad\qquad\qquad\qquad\qquad\qquad\qquad
 \ene\ees
  ii) if we choose other free functions as the periodic functions
 of time $t$, i.e. $\alpha(t)={\rm dn}(t,k), \ \beta(t)={\rm cn}(t,k)$,
 then Figures  2a and 2b display the dynamical behavior of the intensity of the rational-like solution
 (\ref{solu1}) for  different terms $\delta(t)={\rm sn}(t,k),\ {\rm  cn}(t,k)$, respectively,
  in which the coefficients $g(t)$ and $v(x,t)$ in Eq. (\ref{nlsv})  are given by
 \bes\bee
 && g(t)=\frac12\,{\rm cn}(t,k)\,{\rm dn}(t,k)e^{[{\rm sech}(t)-1]/5}, \\
 && v(x,t)=\frac{k^2{\rm cn}(t,k)}{2\,{\rm dn}^2(t,k)}x^2+\frac12\,{\rm cn}(t,k)[k^2{\rm sd}(t,k)x-1]^2 \ \
  {\rm for} \ \  \delta(t)={\rm sn}(t,k), \\
 && v(x,t)=\frac{k^2{\rm cn}(t,k)}{2\,{\rm dn}^2(t,k)}x^2+\frac{{\rm dn}(t,k)}{{\rm
 cn}^2(t,k)}x +\frac12{\rm cn}(t,k)[k^2{\rm sd}(t,k)x+{\rm sc}(t,k)]^2 \ \  {\rm for} \ \  \delta(t)={\rm
 cn}(t,k);
 \ene \ees
 iii) if we choose  other free functions as the periodic  functions
 of time $t$, i.e. $\alpha(t)={\rm cn}(t,k), \ \beta(t)={\rm dn}(t,k)$,
 then Figures 3a and 3b exhibit the dynamical behavior of the rational-like solution
 (\ref{solu1}) for  different terms $\delta(t)={\rm sn}(t,k),\ {\rm  dn}(t,k)$, respectively,
 in which the coefficients $g(t)$ and $v(x,t)$ in Eq. (\ref{nlsv}) are given by \bes\bee
 && g(t)=\frac12\,{\rm cn}(t,k)\,{\rm dn}(t,k)e^{[{\rm sech}(t)-1]/5}, \\
 && v(x,t)=\frac{{\rm dn}(t,k)}{2\,{\rm cn}^2(t,k)}x^2+\frac12\,{\rm dn}(t,k)[{\rm sc}(t,k)x-1]^2 \ \
  {\rm for} \ \  \delta(t)={\rm sn}(t,k), \\
 && v(x,t)=\frac{k^2{\rm dn}(t,k)}{2\,{\rm cn}^2(t,k)}x^2+\frac{k^2{\rm cn}(t,k)}{{\rm
 dn}^2(t,k)}x +\frac12{\rm dn}(t,k)[{\rm sc}(t,k)x+k^2{\rm sd}(t,k)]^2 \ \  {\rm for} \ \  \delta(t)={\rm
 dn}(t,k).
 \ene \ees
 It follows from these figures that the rational-like solution (\ref{solu1}) is
 different from the known rational
 solution of the NLS equation (\ref{nls})~\cite{AB, TP, ABC, ABC1, ABC2,
 ABCP, ABCP2}, and may be useful to raise the possibility  of
 relative experiments and potential applications for the RW
 phenomenon.

\quad {\it Second-order rational-like solution}--- It follows from
(\ref{syst}) that we have the second-order rational solution of
Eq. (\ref{nlsv}) in the form
$P(\eta,\tau)=\mathcal{P}_2(\eta,\tau)/[b\mathcal{A}_2(\eta,\tau)]$
and $
Q(\eta,\tau)=\mathcal{Q}_2(\eta,\tau)/[c\mathcal{A}_2(\eta,\tau)]$,
where
 \bee \label{solu20}
 \begin{array}{l}
 \mathcal{P}_2(\eta(x,t),\tau(t))=\displaystyle -\frac{1}{2}\eta^4-6\eta^2\tau^2-10\tau^4-\frac{3}{2}\eta^2
 -9\tau^2+\frac38,
\vspace{0.1in}\cr \mathcal{Q}_2(\eta(x,t),\tau(t))=\displaystyle
-\tau\left[\eta^4+4\eta^2\tau^2+4\tau^4-3\eta^2+2\tau^2-\frac{15}{4}\right],
\vspace{0.1in}\cr
 \mathcal{A}_2(\eta(x,t),\tau(t))=\displaystyle
 \frac{1}{12}\eta^6+\frac12\eta^4\tau^2+\eta^2\tau^4+\frac23\tau^6
  +\frac18\eta^4
  +\frac92\tau^4-\frac32\eta^2\tau^2+\frac{9}{16}\eta^2+\frac{33}{8}\tau^2+\frac{3}{32},\end{array} \ene
As a consequence, based on the transformations (\ref{comtran}) and
(\ref{newtran}), we have the second-order rational-like solution
(two-rogon solution) of Eq. (\ref{nlsv})
 \bee \label{solu2}
 \psi_2(x,t)=a_0\sqrt{|\alpha(t)|}\,e^{\int^t_0\gamma(s)ds}
  \left[1+\frac{\mathcal{P}_2(\eta(x,t),\tau(t))+i\mathcal{Q}_2(\eta(x,t),\tau(t))}
   {\mathcal{A}_2(\eta(x,t),\tau(t))}\right]e^{i[\chi(x,t)+\tau(t)]} \qquad\qquad
  \ene
whose intensity is given by
 \bee \label{inten2}
 |\psi_2(x,t)|^2=a_0^2|\alpha(t)|\,e^{2\int^t_0\gamma(s) ds}
 \frac{[\mathcal{A}_2(\eta(x,t),\tau(t))+\mathcal{P}_2(\eta(x,t),\tau(t))]^2+\mathcal{Q}_2^2(\eta(x,t),\tau(t))}
  {\mathcal{A}_2^2(\eta(x,t),\tau(t))},  \quad \ene
where $\eta(x,t)=\alpha(t)x+\delta(t)$, and $\chi(x,t),\ \tau(t)$
are given by Eqs. (\ref{simis1}) and (\ref{solu3}). The obtained
second-order rational-like solution (\ref{solu2}) is also
different from the known rational solutions of the NLS equation
(\ref{nls})~\cite{ABC,ABCP}, since some free functions of time $t$
are involved. Similar to the first-order rational-like solution,
the solution (\ref{solu2}) can also reduce to the known one of Eq.
(\ref{nls}) in Refs.~\cite{ABC,ABCP}.

 \quad For the fixed parameters $\alpha_0=1,\ \chi_0(t)=0,\ \gamma(t)=0.1\tanh(t){\rm sech}(t)$,
 i) if we choose other free functions as the polynomials of time $t$, i.e. $\alpha(t)=1, \ \beta(t)=0.5t^2$,
 then Figures 4a and 4b depict the dynamical interaction of the two-rogon solution (\ref{solu2})
 for different terms $\delta(t)=0.1t,\ t^2$; ii) if we choose the free functions as the periodic functions
 of time $t$, i.e. $\alpha(t)={\rm dn}(t,k), \ \beta(t)={\rm cn}(t,k)$,
 then Figures 5a and 5b illustrate the dynamical interaction of the two-rogon
 solution (\ref{solu2}) for  different terms $\delta(t)={\rm sn}(t,k),\ {\rm  cn}(t,k)$;
 iii) if we take other free functions as the periodic functions of time $t$, i.e.
 $\alpha(t)={\rm cn}(t,k), \ \beta(t)={\rm dn}(t,k)$, then Figures 6a and 6b exhibit the dynamical interaction of
 the two-rogon solution (\ref{solu2}) for  different terms $\delta(t)={\rm sn}(t,k),\ {\rm dn}(t,k)$.

\quad In conclusion, we have presented the analytical first-order
and second-order rational-like solution pairs (nonautonomous
rogons) of the inhomogeneous nonlinear Schr\"odinger equation with
variable coefficients by using the similarity transformation and
direct ansatz. By using the direct approach~\cite{ABCP}, we can
also obtain the higher-order rational-like solutions of Eq.
(\ref{nlsv}) which are omitted here. These obtained solutions be
used to describe the possible formation mechanisms for the optical
rogue wave phenomenon in optical fibres, the oceanic rogue wave
phenomenon in the deep ocean, and the matter rogue wave phenomenon
in Bose-Einstein condensates. Furthermore, it should be emphasized
that we give the explicit solutions for the existence of matter
rogons in Bose-Einstein condensates. Moreover, the snake
propagation traces and the interaction of optical nonautonomous
rogons are exhibited by choosing some free functions of time. Some
shapes of the optical nonautonomous rogons and fascinating
interactions between two optical nonautonomous rogons are also
achieved with different functions. These solutions modify the
known solutions related to the optical rogons. Moreover, this
constructive idea can be also extended to other nonlinear systems
with variable coefficients to generate nonautonomous rogons. This
will further excite the study of rogons in the field of nonlinear
science.

 \vspace{0.1in}

\noindent {\bf Acknowledgements}

\vspace{0.05in}

This work was supported by the FCT SFRH/BPD/41367/2007 and the
NSFC60821002/F02.}



\newpage
$~~$ \vspace{-0.0in}

\baselineskip=18pt

\centerline{\Large \bf List of the Figure Captions}

\vspace{0.2in} \noindent
\begin{enumerate}

 \item[Figure 1.]   Wave propagations (left
column ) and contour plots (right column) for the intensity
$|\psi_1|^2$ (\ref{inten1}) of the first-order rational-like
solution (\ref{solu1} for $\alpha=\alpha_0=1.0, \ \beta=0.5t^2, \
\gamma(t)=0.1\tanh(t){\rm sech}(t)$. (a)-(b) $\delta(t)=t$;
(c)-(d) $\delta(t)=t^2$.

\item[Figure 2.] Wave propagations (left column) and contour plots
(right column) for the intensity $|\psi_1|^2$ (\ref{inten1}) of
the first-order rational-like solution (\ref{solu1}) for
$\alpha_0=1.0,\ \gamma(t)=0.1\tanh(t){\rm sech}(t), \ k=0.6,\
\alpha={\rm dn}(t,k), \ \beta(t)={\rm cn}(t,k)$: (a)-(b)
$\delta(t)={\rm sn}(t,k)$; (c)-(d) $\delta(t)={\rm cn}(t,k)$.

 \item[Figure 3.]  Wave propagations (left
column) and contour plots (right column) for the intensity
$|\psi_1|^2$ (\ref{inten1}) of the first-order rational-like
solution (\ref{solu1}) for $\alpha_0=1.0,\
\gamma(t)=0.1\tanh(t){\rm sech}(t), \ k=0.6,\ \alpha={\rm
cn}(t,k), \ \beta(t)={\rm dn}(t,k)$: (a)-(b) $\delta(t)={\rm
sn}(t,k)$; (c)-(d) $\delta(t)={\rm dn}(t,k)$.

 \item[Figure 4.]  Wave propagations (left
column) and contour plots (right column) for the intensity
$|\psi_2|^2$ (\ref{inten2})of the second-order rational-like
solution (\ref{solu2}) for $\alpha_0=1.0,\
\gamma(t)=0.1\tanh(t){\rm sech}(t), \ k=0.6,\ \alpha={\rm
dn}(t,k), \ \beta(t)={\rm cn}(t,k)$: (a)-(b) $\delta(t)=0.1t$;
(c)-(d) $\delta(t)=t^2$.

 \item[Figure 5.] Wave propagations (left
column) and contour plots (right column) for the intensity
$|\psi_2|^2$ (\ref{inten2}) of the second-order rational-like
solution (\ref{solu2}) for $\alpha_0=1.0,\
\gamma(t)=0.1\tanh(t){\rm sech}(t), \ k=0.6,\ \alpha={\rm
dn}(t,k), \ \beta(t)={\rm cn}(t,k)$: (a)-(b) $\delta(t)={\rm
sn}(t,k)$; (c)-(d) $\delta(t)={\rm cn}(t,k)$.

  \item[Figure 6.] Wave propagations (left
column) and contour plots (right column) for the intensity
$|\psi_2|^2$ (\ref{inten2}) of the second-order rational-like
solution (\ref{solu2}) for $\alpha_0=1.0,\
\gamma(t)=0.1\tanh(t){\rm sech}(t), \ k=0.6,\ \alpha={\rm
cn}(t,k), \ \beta(t)={\rm dn}(t,k)$: (a)-(b) $\delta(t)={\rm
sn}(t,k)$; (c)-(d) $\delta(t)={\rm dn}(t,k)$.

\end{enumerate}

\newpage

\begin{figure}
\begin{center}
\vspace{-0.3in}
{\scalebox{0.65}[0.52]{\includegraphics{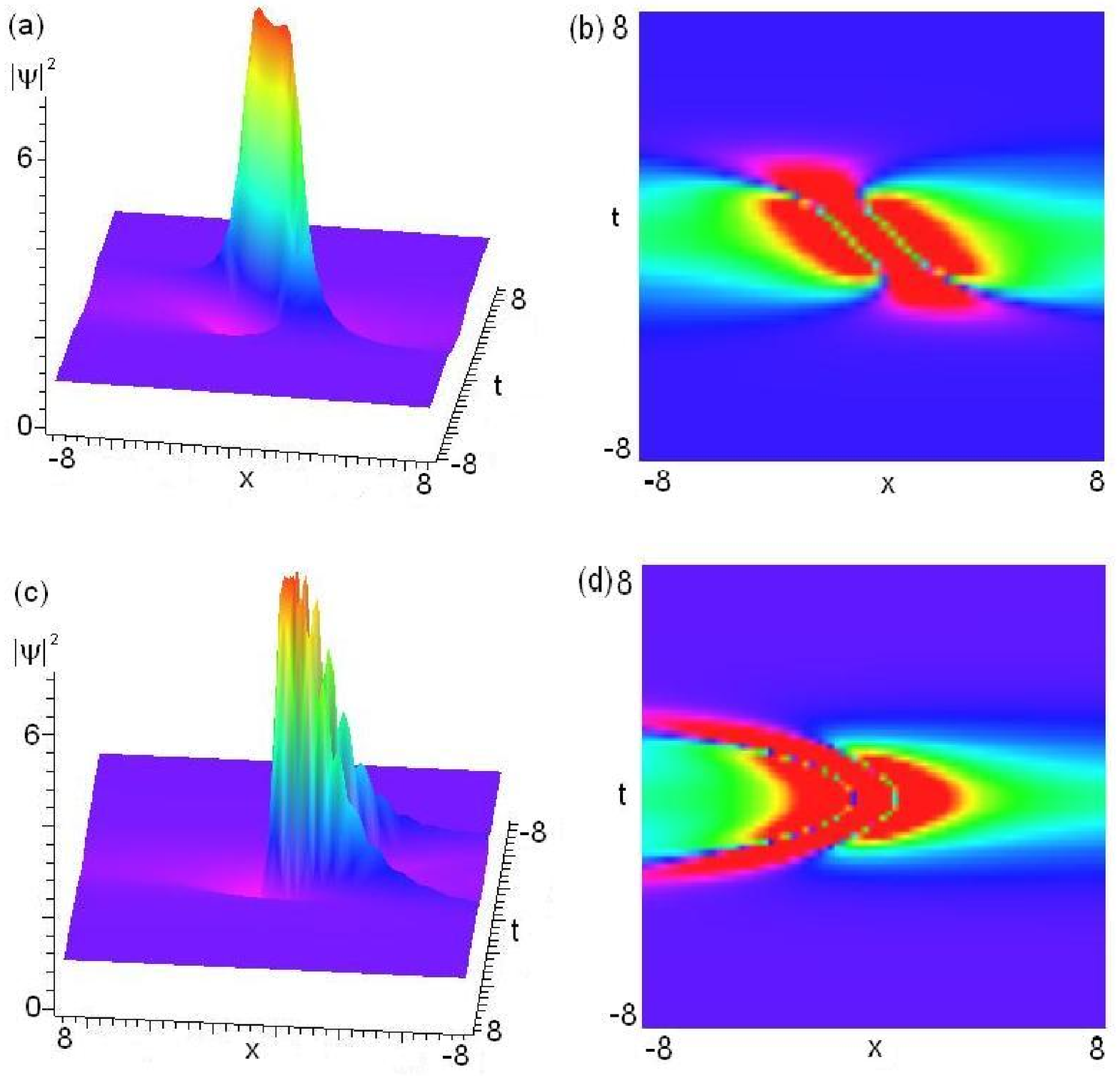}}}
\end{center}
\vspace{-0.25in} \caption{\small (color online). Wave propagations
(left column) and contour plots (right column) for the intensity
$|\psi_1|^2$ (\ref{inten1}) of the first-order rational-like
solution (\ref{solu1}) for $\alpha=\alpha_0=1.0, \ \beta=0.5t^2, \
\gamma(t)=0.1\tanh(t){\rm sech}(t)$. (a)-(b) $\delta(t)=t$;
(c)-(d) $\delta(t)=t^2$. }
\end{figure}

\begin{figure}
\begin{center}
{\scalebox{0.65}[0.52]{\includegraphics{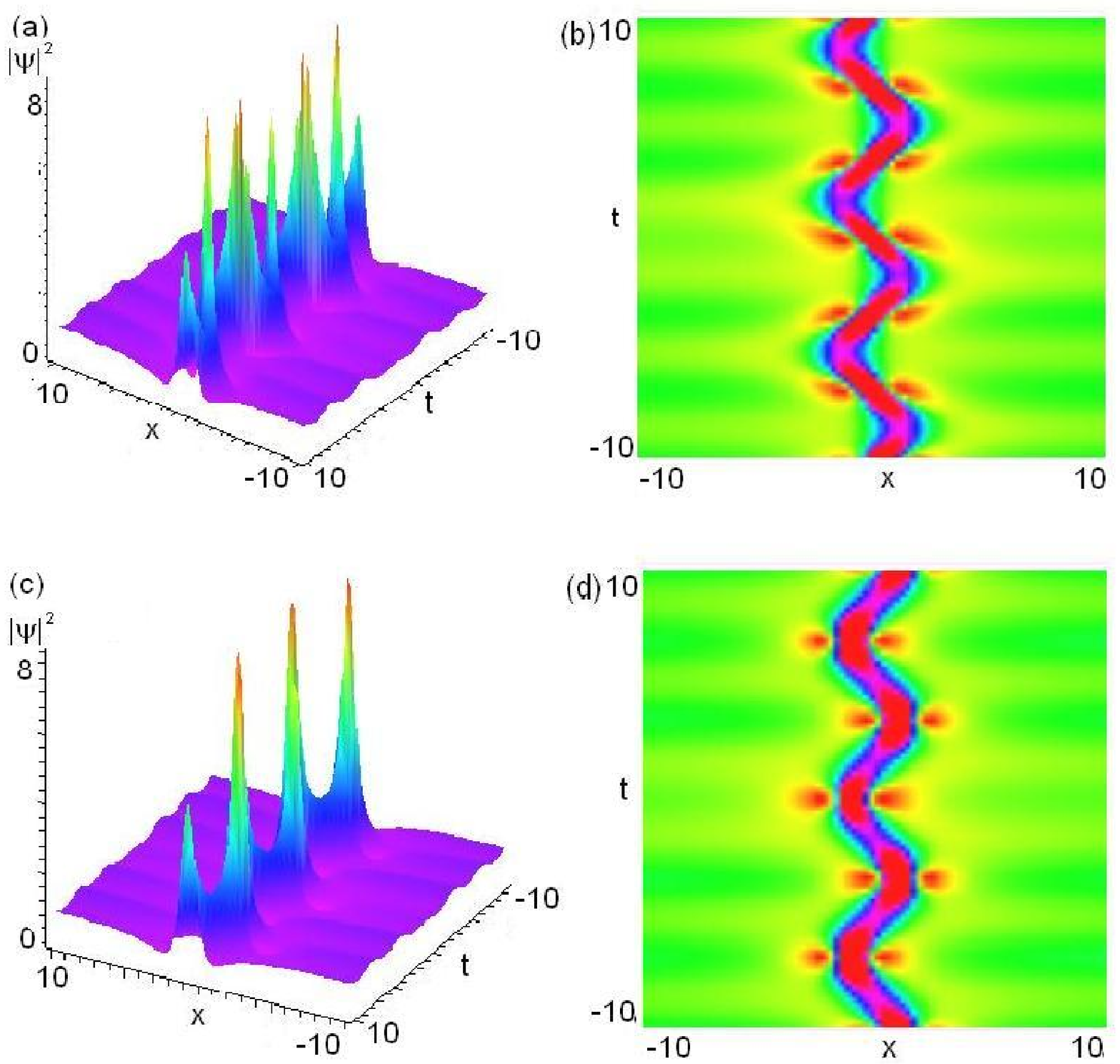}}}
\end{center}
\vspace{-0.25in} \caption{\small (color online). Wave propagations
(left column) and contour plots (right column) for the intensity
$|\psi_1|^2$ (\ref{inten1}) of the first-order rational-like
solution (\ref{solu1}) for $\alpha_0=1.0,\
\gamma(t)=0.1\tanh(t){\rm sech}(t), \ k=0.6,\ \alpha(t)={\rm
dn}(t,k), \ \beta(t)={\rm cn}(t,k)$: (a)-(b) $\delta(t)={\rm
sn}(t,k)$; (c)-(d) $\delta(t)={\rm cn}(t,k)$. }
\end{figure}


\newpage

\begin{figure}
\begin{center}
\vspace{-0.3in}
{\scalebox{0.65}[0.53]{\includegraphics{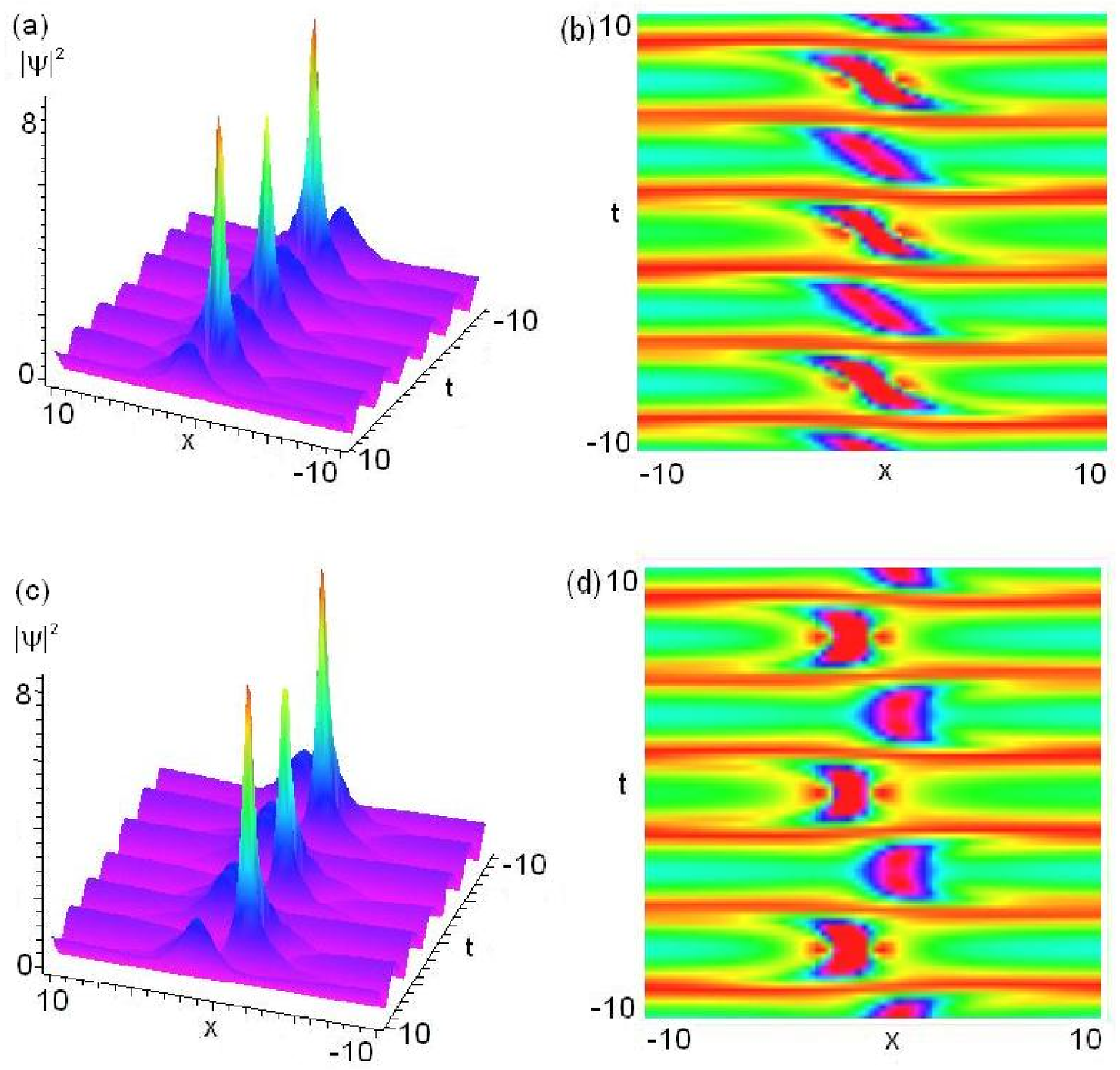}}}
\end{center}
\vspace{-0.25in} \caption{\small (color online). Wave propagations
(left column) and contour plots (right column) for the intensity
$|\psi_1|^2$ (\ref{inten1}) of the first-order rational-like
solution (\ref{solu1}) for $\alpha_0=1.0,\
\gamma(t)=0.1\tanh(t){\rm sech}(t), \ k=0.6,\ \alpha(t)={\rm
cn}(t,k), \ \beta(t)={\rm dn}(t,k)$: (a)-(b) $\delta(t)={\rm
sn}(t,k)$; (c)-(d) $\delta(t)={\rm dn}(t,k)$.}
\end{figure}

\begin{figure}
\begin{center}
{\scalebox{0.65}[0.53]{\includegraphics{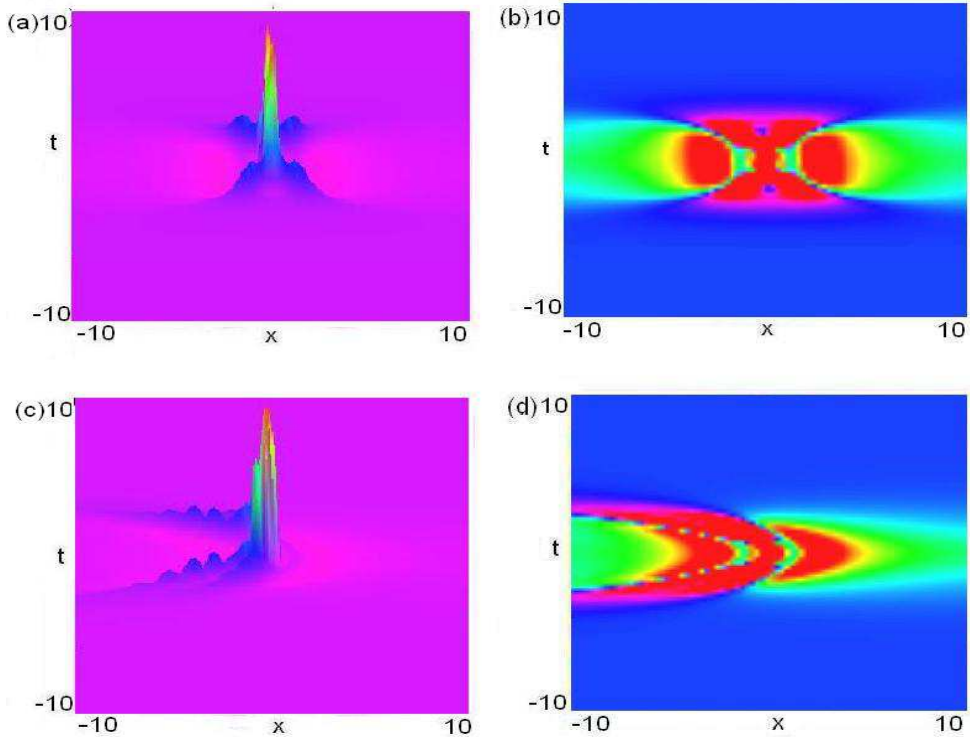}}}
\end{center}
\vspace{-0.25in} \caption{\small (color online). Wave propagations
(left column) and contour plots (right column) for the intensity
$|\psi_2|^2$ (\ref{inten2}) of the second-order rational-like
solution (\ref{solu2}) for $\alpha=\alpha_0=1.0,\
\gamma(t)=0.1\tanh(t){\rm sech}(t), \ \beta(t)=0.5t^2$: (a)-(b)
$\delta(t)=0.1t$; (c)-(d) $\delta(t)=t^2$. }
\end{figure}


\newpage

\begin{figure}
\begin{center}
\vspace{-0.3in}{\scalebox{0.65}[0.55]{\includegraphics{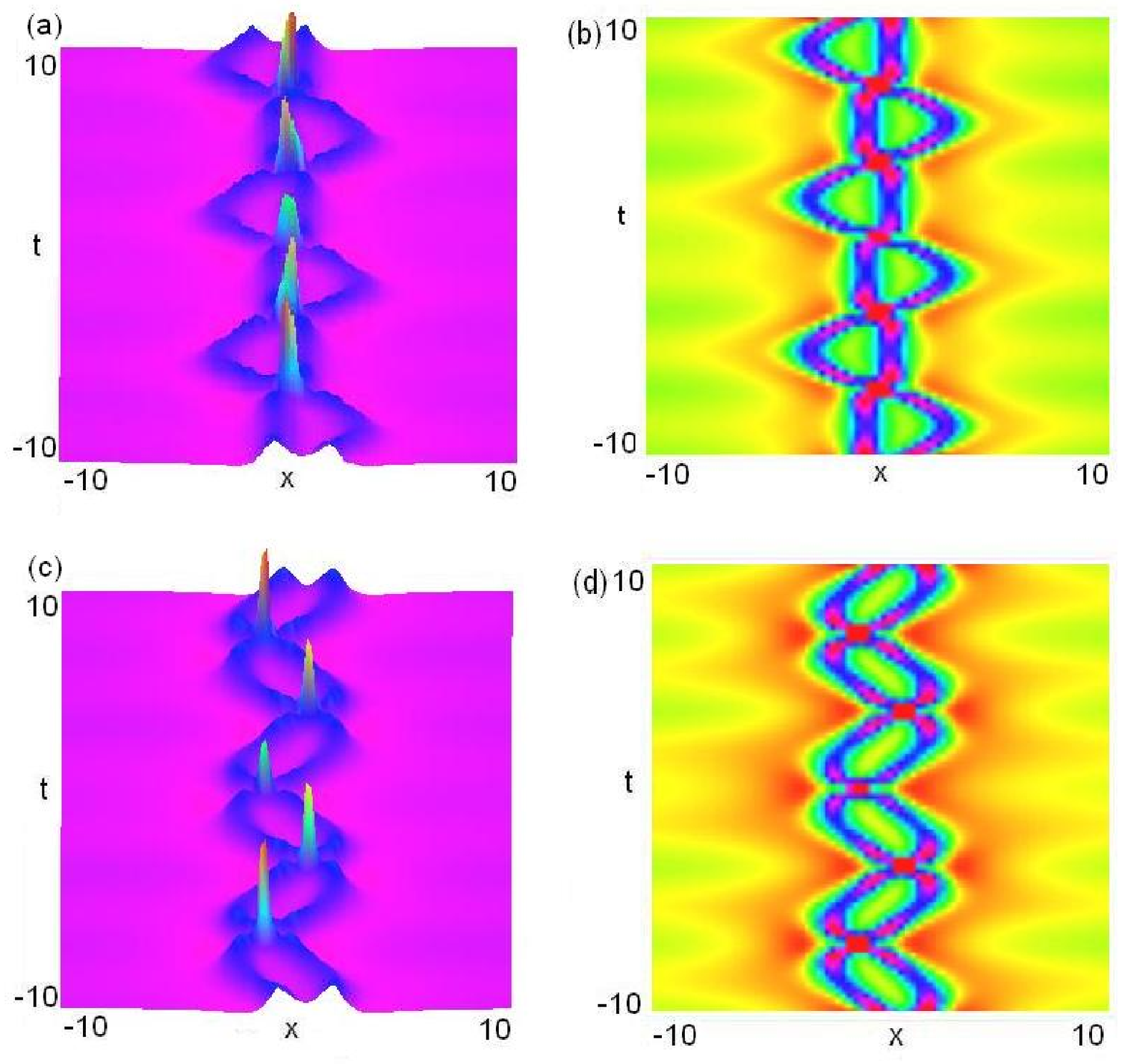}}}
\end{center}
\vspace{-0.25in} \caption{\small (color online). Wave propagations
(left column) and contour plots (right column) for the intensity
$|\psi_2|^2$ (\ref{inten2}) of the second-order rational-like
solution (\ref{solu2}) for $\alpha_0=1.0,\
\gamma(t)=0.1\tanh(t){\rm sech}(t), \ k=0.6,\ \alpha(t)={\rm
dn}(t,k), \ \beta(t)={\rm cn}(t,k)$: (a)-(b) $\delta(t)={\rm
sn}(t,k)$; (c)-(d) $\delta(t)={\rm cn}(t,k)$. }
\end{figure}


\newpage

\begin{figure}
\begin{center}
{\scalebox{0.65}[0.55]{\includegraphics{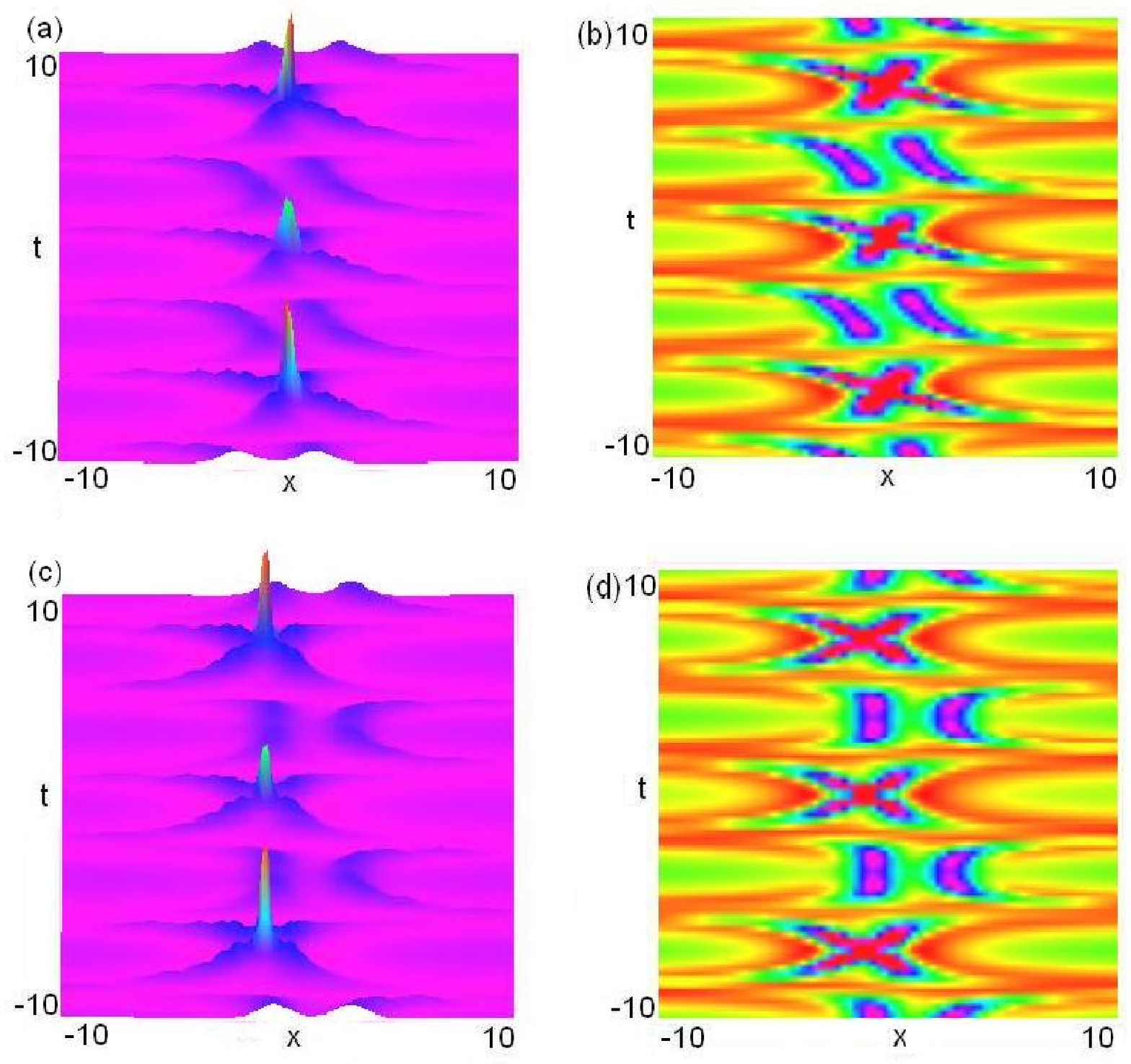}}}
\end{center}
\vspace{-0.25in} \caption{\small (color online). Wave propagations
(left column) and contour plots (right column) for the intensity
$|\psi_2|^2$ (\ref{inten2}) of the second-order rational-like
solution (\ref{solu2}) for $\alpha_0=1.0,\
\gamma(t)=0.1\tanh(t){\rm sech}(t), \ k=0.6,\ \alpha(t)={\rm
cn}(t,k), \ \beta(t)={\rm dn}(t,k)$: (a)-(b) $\delta(t)={\rm
sn}(t,k)$; (c)-(d) $\delta(t)={\rm dn}(t,k)$.}
\end{figure}

\end{document}